\documentclass[rsi,preprint]{revtex4-1}

\usepackage{graphicx}
\usepackage{dcolumn}
\usepackage{bm}
\usepackage{color}
\usepackage{soul}

\draft 

\begin{document}


\title{Synchronization-based image reconstruction for three-dimensional wide-field confocal imaging of periodically moving objects beyond the frame rate} 



\author{Kanta Adachi}
\email{adachi@mech.eng.osaka-u.ac.jp}
\affiliation{Graduate School of Engineering, The University of Osaka, 2-1 Yamadaoka, Suita, Osaka 565-0871, Japan}

\author{Soyoka Hemmi}
\affiliation{Graduate School of Engineering, The University of Osaka, 2-1 Yamadaoka, Suita, Osaka 565-0871, Japan}

\author{Nobutomo Nakamura}
\affiliation{Graduate School of Engineering, The University of Osaka, 2-1 Yamadaoka, Suita, Osaka 565-0871, Japan}


\date{\today}

\begin{abstract}
We extend our previously proposed image reconstruction method, which allows confocal microscopes to capture periodically moving objects at frequencies beyond their frame rates, to three-dimensional and two-dimensional wide-field imaging. This extension is achieved by implementing a synchronization scheme between a confocal laser scanning microscope and a function generator to ensure consistent initial phase alignment across image sequences acquired at different focal depths or fields of view. The method was demonstrated by visualizing the three-dimensional motion of silica particles attached to an aluminum bar oscillating at 100 Hz and the two-dimensional wide-field response of colloidal particles subjected to periodic pulsed excitation. Quantitative single-particle analysis confirmed that the reconstructed images accurately captured the underlying particle dynamics. The extended approach requires no additional specialized hardware and can be readily integrated with conventional confocal microscopes. Thus, it extends the applicability of confocal imaging to the fast dynamics of periodic processes in biological and soft-matter systems.
\end{abstract}

\pacs{}

\maketitle 

\section{Introduction}
Confocal microscopy is widely used in biological and soft-matter research because it enables high-resolution imaging deep within samples and optical sectioning for three-dimensional reconstruction \cite{Schall2004,Prasad2007,Jenkins2008,Elliott2020,Jonkman2020,Kim2024,Terdik2024,Terdik2025}. Its basic operating principle is to focus the excitation light onto a diffraction-limited point in the sample and to reject out-of-focus light using a pinhole placed in front of the detector. By scanning the excitation beam point by point across the sample, confocal microscopy yields optical sections with low background and minimal out-of-focus interference. High-resolution three-dimensional images are obtained by stacking the optical sections acquired in the axial direction.

However, confocal microscopy has a fundamental limitation: pixel-by-pixel image acquisition restricts the achievable frame rate, making a high imaging speed a long-standing technical challenge \cite{Xiao1988,Winter2014,Straub2020}. In this context, confocal microscopes can be broadly categorized into single-point, multipoint, and hybrid scanning microscopes \cite{Elliott2020}. Single-point scanning confocal microscopes, including conventional laser scanning confocal microscopes, are the most common commercial implementation. In these systems, the frame rate is determined directly by the rate of point-by-point data acquisition. Replacing the galvanometer mirrors in these systems with resonant scanning mirrors or acousto-optic deflectors increases the single-beam scan rate, thereby improving imaging speed \cite{Callamaras1999,Im2005,Leybaert2005,Bansal2006}. Multipoint scanning confocal microscopes, most notably spinning-disk confocal microscopes \cite{Petran1968,Tanaami2002,Shimozawa2013,Rodrigo2026}, achieve higher frame rates by scanning multiple laser beams simultaneously and collecting fluorescence from multiple focal points in parallel \cite{Watson2017}. Hybrid scanning confocal microscopes adopt an intermediate approach between single-point and multipoint scanning and use slit apertures rather than pinholes to reject out-of-focus light \cite{Sheppard1988,Wolleschensky2006,Castellano2012,Chu2026}. To date, the highest reported frame rates in confocal microscopy have been achieved using this approach, reaching up to 16,000 frames per second for image sizes of 40 {\textmu}m $\times$ 42 {\textmu}m \cite{Mikami2018}. While these strategies substantially enhance imaging speed, they typically involve increased instrumental complexity and compromises in spatial resolution or field of view.

Recently, we developed a technique that enables conventional laser scanning confocal microscopes to image periodically moving objects at frequencies exceeding the nominal frame rate \cite{Adachi2025}. The method overcomes the limitation imposed by slow image acquisition by reconstructing images based on the phase alignment of each scan line with an external oscillation. Since reconstruction requires no additional hardware and preserves intrinsic imaging performance, the approach is compatible with a wide range of confocal microscopes. In its initial implementation, this method was applied to two-dimensional imaging of colloidal particle motion during elastic wave propagation at frequencies exceeding the frame rate of the microscope. In this case, only the relative phase between scan lines was required, so that the reconstructed images were independent of the absolute initial phase of each image sequence. This freedom in choosing the initial phase simplified implementation and was sufficient for planar imaging of harmonic particle motion. Because the method reconstructs images from a confocal image sequence acquired over multiple cycles of a periodic phenomenon, its application is limited to reproducible dynamic processes.

In principle, this method can be extended to two- and three-dimensional wide-field imaging. The present study primarily aimed to demonstrate these extensions by visualizing the three-dimensional motion of particles attached to an oscillating bar at frequencies exceeding the frame rate and the two-dimensional nonharmonic motion of colloidal particles during pulsed wave propagation across multiple fields of view. To achieve this, we implemented a synchronization scheme between a confocal laser scanning microscope and a function generator. Synchronization ensured that laser scanning was initiated at identical times across successive image sequences, allowing reconstructed images acquired at different focal depths to be stacked into a volumetric dataset and images corresponding to the same phase of motion obtained in successive fields of view to be stitched together.

\section{Experimental details}
An aluminum bar with silica particles (diameter: 1.5 {\textmu}m) that adhered strongly to its tip as a result of our previous colloidal oscillation experiments \cite{Nakamura2017,Nakamura2019,Adachi2025} was mounted on a piezo stage. For confocal imaging of the adhered particles, the aluminum bar was immersed in a dimethyl sulfoxide/water mixture containing a fluorescein sodium salt solution. The mixture composition was chosen such that its refractive index closely matched that of the silica particles. A coordinate system was defined with the \textit{x}-axis parallel to the bar’s vibration direction, the \textit{z}-axis along the depth, and the \textit{y}-axis orthogonal to the \textit{x}- and \textit{z}-axes. The three-dimensional motion of the attached particles was observed by acquiring a confocal image stack of 30 slices while applying a 100 Hz harmonic vibration to the aluminum bar. Confocal images (512 $\times$ 512 pixels) were acquired using a Nikon A1R confocal laser scanning microscope at 15 frames per second. At each focal plane, 55 raw images were acquired under mechanical vibration, after which the stage was moved to the next focal plane. To match the effective spatial resolution in the lateral and axial directions, the resulting 30-layer image stack was linearly interpolated into an 80-layer stack in ImageJ \cite{Schneider2012}.

For the two-dimensional wide-field imaging experiment, a colloidal suspension with a 2\% particle volume fraction was prepared by dispersing the silica particles in the same solvent used in the three-dimensional imaging experiment. To suppress structural changes in the colloid sample during confocal imaging across multiple fields of view under periodic external forcing, poly(sodium 4-styrenesulfonate) with a concentration of 10 {\textmu}mol/L was added to the suspension to induce an attractive depletion interaction between particles \cite{Asakura1958,Nakamura2021}. Thereafter, 600 {\textmu}L of the suspension was placed in a cylindrical aluminum cell (inside diameter: 13 mm) with a 0.15-mm-thick glass coverslip attached to the bottom. The sample was then centrifuged at 1000 rpm for 10 min to form colloidal glass on the coverslip. Subsequently, an aluminum bar mounted on a piezo stage was carefully inserted into the glass, with its tip positioned 20 {\textmu}m above the coverslip, and the sample was left undisturbed for 24 h to stabilize the colloidal structure. Thereafter, periodic pulsed excitation with a period of 0.4 s and a pulse width of 4 ms was applied via the aluminum bar. Confocal images were acquired under these conditions with the same confocal laser scanning microscope used in the earlier experiment. A sequence of confocal images was obtained at a depth of 40 {\textmu}m above the coverslip, with the \textit{x}-position systematically varied in steps of 50 {\textmu}m. 

\begin{figure}
\includegraphics{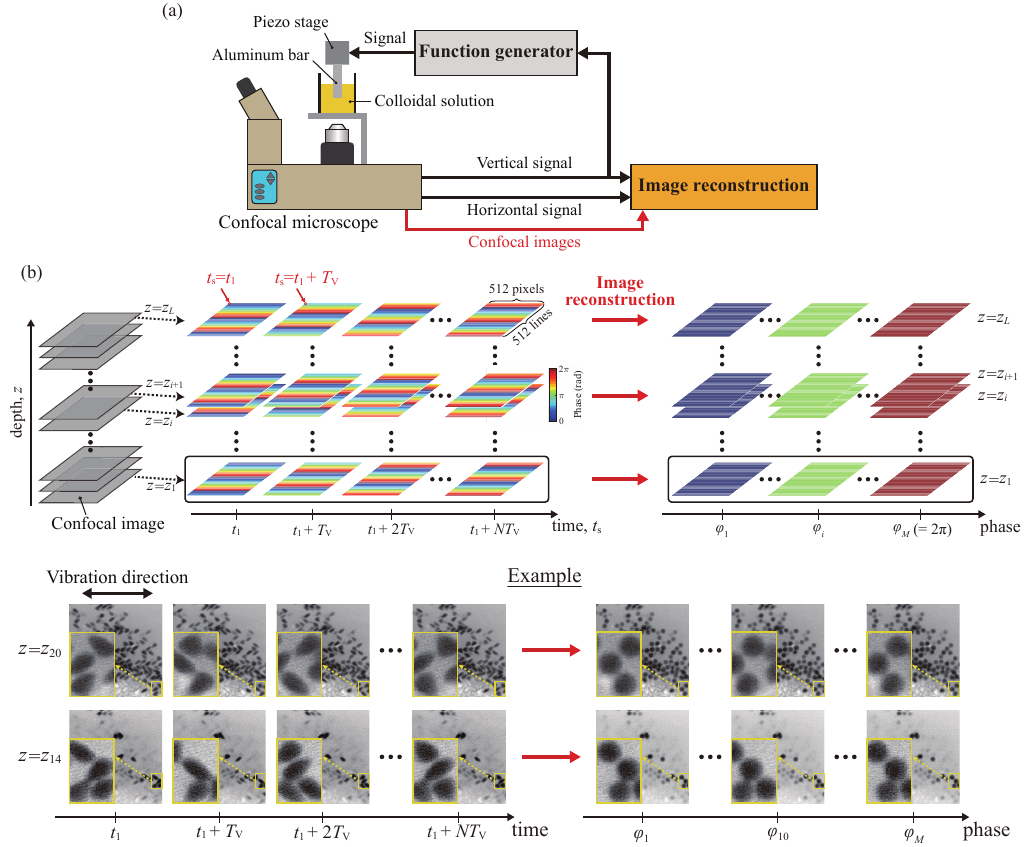}
\caption{(a) Block diagram of a confocal microscopy analysis system. The confocal microscope and function generator are synchronized to ensure a consistent absolute initial phase for each image sequence. (b) Schematic illustration of the image reconstruction process for the three-dimensional imaging of objects moving at frequencies exceeding the frame rate. Pixel size is not to scale. Representative raw and reconstructed confocal images of particles adhered to an aluminum bar oscillating at 100 Hz, acquired at two depths, are also shown.}
\label{Figure_SchematicOverview}
\end{figure}

Figure \ref{Figure_SchematicOverview}(a) shows a block diagram of the measurement system. The vertical and horizontal synchronization signals of the microscope were recorded to determine the acquisition time intervals for the confocal image, $T_{\rm{V}}$, and scan line, $T_{\rm{H}}$. The relative phase, $\varphi '(p, q)$, of each scan line with respect to the periodic phenomenon of interest used for confocal image reconstruction is given by
\begin{equation}
\varphi '(p, q) = 2\pi f \{(p-1)\times T_{\rm{V}}+(q-1)\times T_{\rm{H}}\}+\varphi_{0},
\end{equation}
where $p$ is the order of a confocal image being acquired, and $q$ is the order of the relevant scan line within that image. For example, when reconstructing $N$ sequential confocal images of size 512 $\times$ 512 pixels, $p=1, 2, \cdots , N$ and $q=1, 2, \cdots , 512$. In two-dimensional imaging within a single field of view, image reconstruction requires only the relative phase between scan lines, and therefore, the initial phase $\varphi_{0}$ is arbitrary. Accordingly, $\varphi_{0}$ was set to zero for convenience in our previous study \cite{Adachi2025}, which demonstrated the initial implementation of this method. However, when extending the method to three-dimensional and two-dimensional wide-field imaging, $\varphi_{0}$ must be identical across different focal planes and fields of view to obtain internally consistent image stacks or wide-field images. Hence, the vertical synchronization signal was used as a trigger for a function generator, thereby synchronizing signal generation with the start of image scanning by the confocal microscope. This synchronization ensures that the $\varphi_{0}$ value remains identical for all confocal image sequences, allowing reconstructed images acquired at different focal planes or fields of view to be combined consistently. 

Figure \ref{Figure_SchematicOverview}(b) schematically illustrates the three-dimensional imaging of objects moving at frequencies exceeding the frame rate. The process begins by acquiring $N$ sequential confocal images at an initial depth ($z=z_{1}$). Image acquisition is then repeated while systematically varying the focal depth, yielding a series of image sequences over the desired depth range ($z=z_{1}, z_{2}, \cdots , z_{L}$). When objects move at frequencies beyond the frame rate, the raw confocal images are generally distorted severely. At each depth, $N$ raw images are reconstructed into $M$ confocal images to correct these distortions. Stacking the reconstructed images according to the same phase of motion yields an accurate three-dimensional representation of objects moving at frequencies exceeding the frame rate. Figure \ref{Figure_SchematicOverview}(b) shows representative confocal image sequences of silica particles adhered to the tip of an aluminum bar oscillating at 100 Hz, acquired at two different depths. While the raw images exhibit pronounced distortion due to the limited frame rate, the reconstructed images clearly resolve the particle positions at each phase of motion. The same reconstruction scheme can be applied to two-dimensional wide-field imaging by laterally shifting the field of view instead of varying the focal depth. Thus, the proposed synchronization scheme enables conventional laser scanning confocal microscopes to perform three-dimensional and two-dimensional wide-field imaging of periodically moving objects at frequencies beyond their nominal frame rates.

\section{Results and discussions}
\begin{figure}
\includegraphics{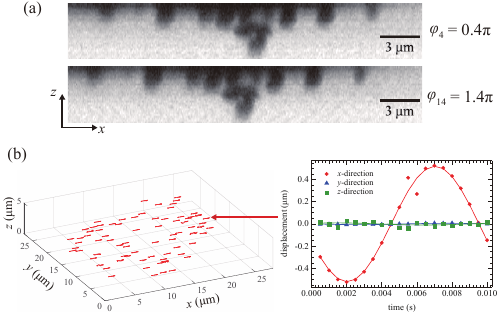}
\caption{(a) \textit{x}--\textit{z} cross-sectional images at phases 0.4$\pi$ and 1.4$\pi$, corresponding to the phases at which the particle displacement along the \textit{x}-axis is maximum. (b) Three-dimensional trajectories of 76 particles over one oscillation period. The displacements of a representative particle along the \textit{x}-, \textit{y}-, and \textit{z}-axes are also shown. Solid lines indicate discrete Fourier transform fits. Multimedia available online.}
\label{Figure_3D_Observation}
\end{figure}

Figure \ref{Figure_3D_Observation}(a) (Multimedia views) shows representative contrast-enhanced \textit{x}--\textit{z} cross-sectional images extracted from the reconstructed 80-layer image stack at phases of 0.4$\pi$ and 1.4$\pi$. In all, 55 raw images were used to reconstruct 20 images at each focal depth ($N=55$ and $M=20$), yielding a sequence of reconstructed images that represents particle motion at regular phase intervals over one oscillation period. Because the axial resolution of confocal microscopy is inherently lower than its lateral resolution \cite{Elliott2020}, the particles appear ellipsoidal rather than circular in the \textit{x}--\textit{z} cross-sectional views (Fig. \ref{Figure_3D_Observation}(a)). Although reducing the pinhole diameter can improve axial resolution and achieve particle shapes closer to circular, the three-dimensional images obtained in the present study provide sufficient spatial resolution for the quantitative analysis of the three-dimensional motion of particles adhered to the aluminum bar (see multimedia \ref{Figure_3D_Observation}). Individual particle positions were tracked using particle-tracking software \cite{Gao2009,Jensen2016}. The resulting trajectories were approximately linear, indicating that the particles oscillated predominantly in a single direction (Fig. \ref{Figure_3D_Observation}(b)). The oscillation amplitude of each particle was quantified by performing a discrete Fourier transform of the displacement time series and isolating the component at the oscillation frequency of the aluminum bar. For a representative particle, the fitted displacement shows that the oscillation amplitude along the \textit{x}-axis is substantially larger than that along the other axes, confirming the harmonic vibration parallel to the \textit{x}-axis at 100 Hz (Fig. \ref{Figure_3D_Observation}(b)). All 76 particles that could be tracked during one oscillation exhibited similar behavior, demonstrating that the particle motion faithfully follows the motion of the aluminum bar. These results confirm that the proposed synchronization scheme enables confocal laser scanning microscopes to capture three-dimensional particle motion at frequencies exceeding the nominal frame rate.

\begin{figure}
\includegraphics{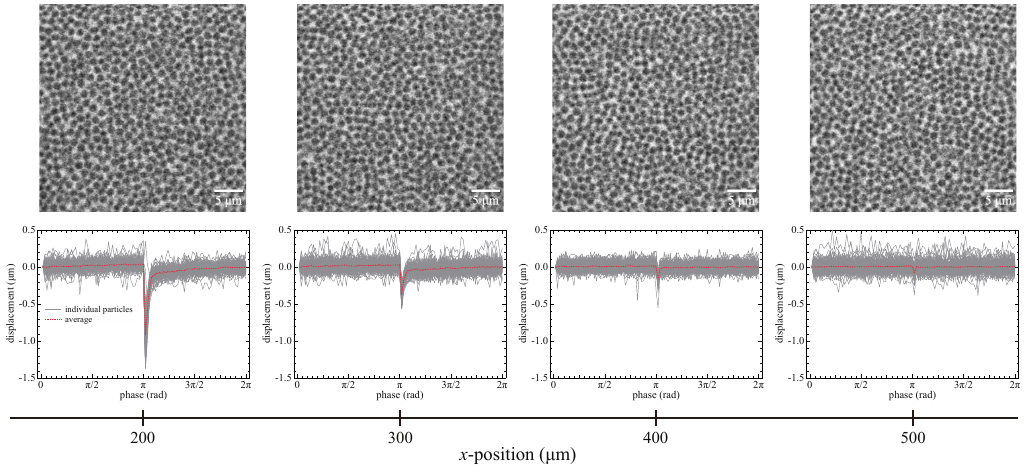}
\caption{Displacement of individual particles along the $x$-axis over one pulse-wave generation period at different \textit{x}-positions. Dotted lines represent the average displacement at each phase. Corresponding snapshots at the initial phase are shown above each graph. Multimedia available online.}
\label{Figure_2D_Observation}
\end{figure}

\begin{figure}
\includegraphics{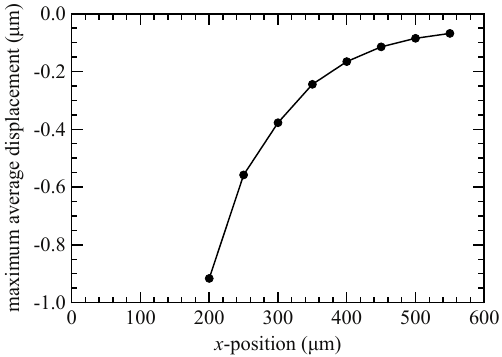}
\caption{Maximum average displacement of colloidal particles as a function of their $x$-position during pulse-wave propagation.}
\label{Figure_attenuation}
\end{figure}

Figure \ref{Figure_2D_Observation} (Multimedia views) shows the particle displacement along the \textit{x}-axis over one pulse-wave generation period at different \textit{x}-positions in a two-dimensional wide-field imaging experiment employing the proposed method. In this experiment, 237 confocal images were acquired at each field of view. After contrast enhancement and mean filtering, the images were reconstructed into 100 images with equally spaced phases within a single pulse-wave cycle ($N=237$ and $M=100$). Individual particle positions were tracked as done in the three-dimensional imaging experiment \cite{Gao2009}, yielding a displacement time series along the pulse-wave propagation direction. The distance-dependent attenuation of pulse waves propagating through the colloidal particles was evaluated using the averaged particle displacement at each \textit{x}-position (Fig. \ref{Figure_attenuation}). However, particle displacements were extremely large near the aluminum bar; accurate reconstruction of the confocal image sequences was not possible in the region $x<200$ {\textmu}m because the colloidal structures were not identical at the beginning and end of each image sequence. Nevertheless, the displacement data demonstrated that the displacement amplitude decreased with increasing distance from the bar (Fig. \ref{Figure_attenuation}), as expected. However, the pulse-wave propagation velocity was substantially higher than the effective frame rate of the reconstructed image sequences, preventing the direct visualization of pulse-wave propagation with spatial attenuation. This indicates that the particle motion associated with the pulse wave effectively propagated instantaneously from the vicinity of the aluminum bar to the furthest field of view (see multimedia \ref{Figure_2D_Observation}). The effective frame rate was determined by $M$. In the present experiment, 100 reconstructed images were used to represent a phenomenon with a period of 0.4 s, corresponding to an effective temporal resolution of 0.004 s. Although this temporal resolution was insufficient to directly resolve the pulse-wave propagation, it could capture the hysteretic behavior of particle displacements between the positive and negative \textit{x}-directions during pulse-wave excitation (Fig. \ref{Figure_2D_Observation}). These results demonstrate that even when direct visualization of pulse-wave propagation is challenging, the mechanical response and nonlinear behavior of colloidal systems can be characterized by analyzing particle motion at each spatial location. Thus, the proposed technique is also effective for two-dimensional wide-field imaging of pulse-wave-induced nonharmonic, fast particle motion using a conventional confocal microscope.

\section{Conclusions}
We implemented a synchronization scheme between a confocal laser scanning microscope and a function generator that extends our previously proposed image reconstruction method to three-dimensional and two-dimensional wide-field imaging. The synchronization ensures consistent phase alignment across image sequences acquired at different focal depths or fields of view, enabling the volumetric reconstruction and wide-area stitching of periodically reproducible dynamics. These extensions preserve the versatility of the original technique without any additional specialized hardware and maintain intrinsic imaging performance. The proposed approach can be readily integrated with conventional confocal microscopes to visualize and quantitatively analyze the fast dynamics of periodic processes in biological and soft-matter systems, including spontaneous and nonharmonic motion.

\section*{author declarations}
\subsection*{Conflict of Interest}
The authors have no conflicts to disclose.

\subsection*{Author Contributions}
\noindent
{\bf Kanta Adachi: }Formal analysis (equal); Investigation (equal); Software (equal); Writing -- original draft (lead); Writing -- Review \& Editing (equal). {\bf Soyoka Hemmi: }Formal analysis (equal); Investigation (equal); Software (equal); Writing -- Review \& Editing (equal). {\bf Nobutomo Nakamura: }Conceptualization (lead); Investigation (equal); Supervision (lead); Writing -- original draft (supporting); Writing -- Review \& Editing (equal).

\section*{data availability}
The data that support the findings of this study are available from the corresponding author upon reasonable request.


%
%

%


\bibliography{mybibfile_20260115_3D_ConfocalMicroscopy}

\clearpage

\end{document}